\documentclass[superscriptaddress,aps,twocolumn,showpacs,nofootinbib]{revtex4}

\usepackage{graphicx}
\usepackage{amsmath,amssymb}

\begin{document}

\title{Brane assisted quintessential inflation with transient acceleration}

\author{M. C. Bento}
\email{bento@sirius.ist.utl.pt}
\affiliation{Centro de F\'{\i}sica Te\'orica de
Part\'{\i}culas, Instituto Superior T\'{e}cnico\\ Avenida Rovisco
Pais, 1049-001 Lisboa, Portugal}

\author{R. Gonz\'{a}lez Felipe}
\email{gonzalez@cftp.ist.utl.pt}
\affiliation{%
Instituto Superior de Engenharia de Lisboa\\ Rua Conselheiro Em\'idio Navarro, 1959-007 Lisboa,
Portugal}
\affiliation{%
Centro de F\'{\i}sica Te\'orica de
Part\'{\i}culas, Instituto Superior T\'{e}cnico\\ Avenida Rovisco
Pais, 1049-001 Lisboa, Portugal}

\author{N.~M.~C. Santos}
\email{ncsantos@cftp.ist.utl.pt}
\affiliation{%
Centro de F\'{\i}sica Te\'orica de
Part\'{\i}culas, Instituto Superior T\'{e}cnico\\ Avenida Rovisco
Pais, 1049-001 Lisboa, Portugal}

\begin{abstract}

A simple model of quintessential inflation with the modified exponential potential $e^{-\alpha \phi} \left[A+\left(\phi-\phi_0\right)^2\right]$ is analyzed in the braneworld context. Considering reheating via instant preheating, it is shown that the evolution of the scalar field $\phi$ from inflation to the present epoch is consistent with the observational constraints in a wide region of the parameter space. The model exhibits transient acceleration at late times for $0.96 \lesssim A \alpha^2 \lesssim 1.26$ and $271 \lesssim \phi_0\, \alpha \lesssim 273$, while permanent acceleration is obtained for $2.3\times10^{-8} \lesssim A \alpha^2 \lesssim 0.98$ and $255 \lesssim \phi_0\, \alpha \lesssim 273$. The steep parameter $\alpha$ is constrained to be in the range $5.3 \lesssim \alpha \lesssim 10.8$.
 \end{abstract}


\date{\today}
\maketitle

\section{Introduction}

The recent measurements of the Wilkinson Microwave Anisotropy Probe (WMAP)~\cite{Bennett:2003bz,Spergel:2003cb,Hinshaw:2006ia,Page:2006hz,Spergel:2006hy,Hinshaw:2008kr,Dunkley:2008ie,Komatsu:2008hk} on the cosmic microwave background (CMB) made it clear that the current state of the Universe is very close to a critical density and that the primordial density perturbations that seeded large-scale structure in the Universe are nearly scale-invariant and Gaussian, which is consistent with the inflationary paradigm. Inflation is often implemented in models with a single or multiple scalar fields~\cite{Lyth:1998xn}, which undergo a slow-roll period allowing an early accelerated expansion of the Universe.

Furthermore, the Universe seems to exhibit an interesting symmetry with regard to the accelerated expansion, namely, it underwent inflation at early epochs and is believed to be accelerating at present. The current acceleration of the Universe is supported by observations of high redshift type Ia supernovae~\cite{Astier:2005qq,Riess:2006fw} and, more indirectly, by observations of the CMB and galaxy clustering~\cite{Spergel:2006hy,Komatsu:2008hk,Seljak:2006bg}. Within the framework of general relativity, cosmic acceleration should be sourced by an energy-momentum tensor which has a large negative pressure (dark energy)~\cite{Sahni:1999gb,Padmanabhan:2002ji,Ratra:1987rm,Wetterich:1987fm,Frieman:1995pm,Ferreira:1997au,Zlatev:1998tr,Brax:1999yv,Barreiro:1999zs,Bento:2001yv}. Therefore, in order to comply with the logical consistency and observations, the standard model should be valid somewhere between inflation at early epochs and quintessence at late times. It is then natural to ask whether one can build a model to join these two ends without disturbing the thermal history of the Universe. Attempts have been made to unify both these concepts using models with a single scalar field~\cite{Peebles:1998qn,Copeland:2000hn,Huey:2001ae,Majumdar:2001mm,Nunes:2002wz,Sami:2004xk,Leith:2007bu,Neupane:2007mu}, i.e., in which a single scalar field plays the role of the inflaton and quintessence - the so-called quintessential inflation.

On the other hand, in recent years there has been increasing interest in the cosmological implications of a certain class of braneworld scenarios where the Friedmann equation is modified at very high energies. In particular, in the Randall-Sundrum type II (RSII) model~\cite{Randall:1999vf} the square of the Hubble parameter, $H^2$, acquires a term quadratic in the energy density, allowing slow-roll inflation to occur for potentials that would be too steep to support inflation in the standard Friedmann-Robertson-Walker (FRW) cosmology~\cite{Cline:1999ts,Kaloper:1999sm,Shiromizu:1999wj,Binetruy:1999ut,Flanagan:1999cu,Maartens:1999hf,Maartens:2003tw,Bento:2006sr}. Indeed, in a cosmological scenario in which the metric projected onto the brane is a spatially flat FRW model, the Friedmann equation in four dimensions reads (after setting the 4D cosmological constant to zero and assuming that inflation rapidly makes any dark radiation term negligible)~\cite{Cline:1999ts}
\begin{equation}
H^2 = {1 \over 3\, M_4^2}\, \rho\, \left[1 + {\rho \over 2 \lambda}\right]~.
\label{eq:Friedmann}
\end{equation}
Here $M_4$ is the 4D reduced Planck mass and $\rho\equiv \rho_{\phi} = \frac{1}{2}{\dot\phi}^2 + V(\phi)$ in a Universe dominated by a single minimally coupled homogeneous scalar field. The brane tension $\lambda$ relates the 4D and 5D Planck masses through the relation
\begin{equation}
\lambda = \frac{3}{32 \pi^2} \frac{M_5^6}{M_4^2}~,
\label{eq:tension}
\end{equation}
where $M_5$ is the 5D Planck mass.

We notice that Eq.~(\ref{eq:Friedmann}) reduces to the usual Friedmann equation at sufficiently low energies, \mbox{$\rho \ll \lambda$}, while at very high energies $H\propto\rho$. In this scenario, all matter fields are confined to the brane and, hence, inflation is driven by a 4D scalar field trapped on the brane with the usual equation of motion
\begin{equation}
{\ddot \phi} + 3H {\dot \phi} + V'(\phi) = 0~,
\label{eq:kg}
\end{equation}
where the prime denotes the derivative with respect to the scalar field $\phi$. From Eqs.~(\ref{eq:Friedmann}) and (\ref{eq:kg}) it becomes clear that the presence of the additional term $\sim \rho^2/\lambda$ increases the damping experienced by the scalar field as it rolls down its potential.

It has been shown that, in the RSII braneworld context, quintessential inflation can occur for a sum of exponentials or cosh potentials~\cite{Majumdar:2001mm,Nunes:2002wz,Sami:2004xk}. In this paper we show that the modified exponential potential (hereafter we adopt natural units, $M_4=1$, unless stated otherwise)
\begin{equation}
V(\phi)=e^{-\alpha \phi} \left[A+\left(\phi-\phi_0\right)^2\right]
\label{eq:albrecht}
\end{equation}
also leads to a successful quintessential inflation model.

In the context of quintessence, this potential was first analyzed by Albrecht and Skordis (AS)~\cite{Albrecht:1999rm}. Afterward, it has been extensively studied in the literature~\cite{Barrow:2000nc,Skordis:2000dz,Barreiro:2003ua,Blais:2004vt,Barnard:2007ta}. Regarding its motivation, it is worth noticing that exponential potentials naturally appear in 4D field theories coming from string/M-theory~\cite{Gasperini:2001pc}, where the scalar field $\phi$ is typically identified with the dilaton field. As far as the origin of the polynomial factor is concerned, it can be associated with a nontrivial K\"ahler term in an effective 4D supergravity theory~\cite{Copeland:2000vh}. The scalar $\phi$ could also be associated with a modulus (radion) field in curled extra dimensions~\cite{Albrecht:2001xt}, which need not be universally coupled to matter/gauge fields and, therefore, is not subject to quantum corrections~\cite{Doran:2002bc}.

The tracking properties of the AS potential are similar to the pure exponential, namely, it allows sufficient radiation domination during big bang nucleosynthesis (BBN), followed by matter domination. Nevertheless, due to the presence of the polynomial factor, the field evolves to quintessence dominance near the present epoch. One should notice that, in order to the transition to happen near the present, the parameter $\phi_0$ must be suitably chosen. In other words, this model does not explain the so-called coincidence problem. However, the model displays an interesting feature: it can lead to both permanent and transient acceleration regimes. Permanent acceleration occurs for $A\alpha^2<1$, when the field is trapped in the local minimum of the potential. Transient vacuum domination arises in two ways~\cite{Barrow:2000nc}: when $A\alpha^2<1$ and the $\phi$ field arrives at the minimum of the potential with enough kinetic energy to roll over the barrier and resumes descending the potential where $\phi \gg \phi_0$, or for $A\alpha^2>1$, when the potential loses its local minimum.

In the models mentioned above, inflation takes place when the pure exponential potential dominates the potential. The exit from inflation takes place naturally when the slow-roll conditions are violated because the high-energy brane corrections become unimportant. Moreover, these models belong to the category of nonoscillating models in which the standard reheating mechanism does not work. In this case, one can employ an alternative mechanism of reheating via  gravitational particle production~\cite{Ford:1986sy,Spokoiny:1993kt,Grishchuk:1990bj}. However, this mechanism is faced with difficulties associated with excessive production of gravity waves. Indeed, the reheating mechanism based upon this process is extremely inefficient. The energy density of the produced radiation is typically one part in $10^{16}$~\cite{Copeland:2000hn} to the scalar field energy density at the end of inflation. As a result, these models have a prolonged kinetic regime during which the amplitude of primordial gravity waves is enhanced and the nucleosynthesis constraints are violated~\cite{Sahni:2001qp}. These problems can be circumvented if one invokes an alternative method of reheating, namely, the so-called instant preheating~\cite{Felder:1998vq,Kofman:1994rk,Kofman:1997yn}. This mechanism is quite efficient and robust, and is well suited to nonoscillating models~\cite{Felder:1999pv}. The larger reheating temperature in this model results in a smaller amplitude of relic gravity waves which is consistent with the BBN bounds~\cite{Sami:2004xk}.

\section{Braneworld inflation with an exponential potential}
\label{sec:inflation}

The exponential potential
\begin{equation}
V(\phi) = V_0\,e^{-\alpha \, \phi}
\label{eq:expo}
\end{equation}
with ${\dot \phi} > 0$ has traditionally played an important role within the inflationary framework  since, in the absence of matter, it gives rise to power-law inflation $a \propto t^{2/\alpha^2}$, provided $\alpha\leq \sqrt{2}$. For $\alpha > \sqrt{2}$ the potential becomes too steep to sustain inflation and for larger values $\alpha \geq \sqrt{6}$ the field enters a kinetic regime during which the field energy density $\rho_\phi \propto a^{-6}$. Thus, within the standard general relativity framework, steep potentials are not capable of sustaining inflation. However, in the RSII scenario, the increased damping of the scalar field when $V/\lambda \gg 1$ leads to a decrease in the value of the slow-roll parameters and inflation becomes possible even for large values of $\alpha$.

The cosmological dynamics with a steep exponential potential in the presence of a background (radiation/matter) admits a scaling solution as the attractor of the system. The attractor is characterized by the tracking behavior of the field energy density $\rho_{\phi}$. During the tracking regime, the ratio of  $\rho_{\phi}$ to the background energy density $\rho_B$ is held fixed,
\begin{equation}
\Omega_\phi \equiv {\rho_{\phi} \over {\rho_{\phi}+\rho_B}}={{3\left (1+w_B \right)} \over \alpha^2}\,,
\label{track}
\end{equation}
where $w_B$ is the equation-of-state parameter for the background ($w_B=0, 1/3$ for matter and radiation, respectively). The field energy density in the post inflationary regime would keep tracking the background being subdominant such that it does not interfere with the thermal history of the Universe. Nevertheless, the polynomial factor in the AS potential will allow the scalar field to dominate near the present time, for a suitable choice of the parameters. If one takes into account the nucleosynthesis constraint $\Omega_\phi^{\rm BBN} \lesssim 0.09$~\cite{Bean:2001wt}, coming from the primordial abundances of $^4$He and D, Eq.~(\ref{track}) would require  $\alpha \gtrsim 6.7$. We notice however that this bound can be slightly relaxed if the scalar field has not yet entered the tracking regime during BBN (see Sec.~\ref{sec:quintessence}).

\subsection{Slow-roll inflation}

Let us first review the slow-roll inflation driven by an exponential potential in the high-energy regime of the RSII braneworld. These results will be important to determine the initial conditions for the solution of the evolution equations at later times.

The number of $e$-folds during the inflationary period is given by~\cite{Maartens:1999hf}
\begin{equation}
N(\phi) = -  \int_{\phi}^{\phi_{\rm end}}\dfrac{ V }{ V'} \left[
1+{ V\over 2\,\lambda}\right] d{\phi}~,\label{eq:Nfolds}
\end{equation}
where $\phi_{\rm end}$ corresponds to the field value at the end of inflation. Braneworld effects at high energies increase the Hubble rate by a factor $V/(2\,\lambda)$, yielding more inflation between any two values of $\phi$ for a given potential. As $V \gg \lambda$ during inflation, one gets
\begin{equation}
N \simeq {1\over 2\lambda \alpha^2}(V_N-V_{\rm end})~,
\label{eq:efold2}
\end{equation}
where $V_N$ is the potential evaluated at $N$ $e$-folds from the end of inflation.

The prediction for the inflationary variables typically depends on the number of $e$-folds of inflation occurring after the observable universe leaves the horizon, $N_\star = N(\phi_\star)$. The calculation of this quantity requires a model of the entire history of the Universe~\cite{Liddle:2003as,Dodelson:2003vq} and, as we shall see later, it can be determined once we set the reheating mechanism after inflation.

Inflation ends when the slow-roll conditions are violated, because the brane high-energy corrections become unimportant. Hence, the value of the potential $V_{\rm end}$ at the end of inflation can be obtained from the condition
\begin{equation}
{\rm max}\{\epsilon(\phi_{\rm end}),|\eta(\phi_{\rm end})|\}= 1~,
\label{eq:phif}
\end{equation}
where the slow-roll parameters are defined as
\begin{align}
\epsilon & = \frac{1}{2} {V^{\prime2} \over V^2} {1+{V/ \lambda}\over(1+{V/ 2
\,\lambda})^2}~,
\label{eq:epsilon}\\
\eta & =  {V'' \over V}   {1 \over 1+{V/ 2\, \lambda}}~. \label{eq:eta}
\end{align}
In the brane high-energy regime, i.e., for $V \gg \lambda$, one obtains
\begin{align}
\epsilon = \eta \simeq \frac{2 \, \alpha^2 \lambda}{V}~,\label{eq:epsiloneta}
\end{align}
leading to
\begin{align}
V_{\rm end} \simeq 2 \, \alpha^2 \lambda~.\label{eq:Vend}
\end{align}
Therefore, taking into account Eq.~(\ref{eq:efold2}), the value of the potential $V_\star$ at horizon crossing is
\begin{equation}
V_\star \simeq V_{\rm end} (N_\star+1).
\label{eq:Vstar}
\end{equation}

In the RSII model, the scalar and tensor perturbation amplitudes are given by \cite{Maartens:1999hf,Langlois:2000ns}
\begin{eqnarray}
A_s^2 &= \dfrac{1}{75 \,\pi^2}\,\dfrac{V^3}
{V^{\prime2}} \left[ 1 + \dfrac{V}{2\,\lambda} \right]^3 ~, \label{eq:As1}\\
A_t^2 &=\dfrac{1}{150 \,\pi^2}\,  V \, \left( 1+\dfrac {V}{ 2\, \lambda}\right)
F^2~, \label{eq:At1}
\end{eqnarray}
where
\begin{equation}
 F^2=\left[\sqrt{1+x^2} -x^2
\sinh^{-1}\left({1\over x}\right)\right]^{-1}~,\label{eq:AtF}
\end{equation}
and
\begin{align}
x\equiv \left(\dfrac{3 \,H^2}{4 \pi\,\lambda}\right)^{1/2} =
\left[\dfrac{2\,V} {\lambda
}\left(1+\dfrac{V}{2\,\lambda}\right)\right]^{1/2}~.\label{eq:Ats}
\end{align}
In the low-energy limit ($x\ll 1$), $F^2\approx 1$, whereas $F^2\approx 3\,
V/2\,\lambda$ in the high-energy limit. The right-hand sides of
Eqs.~(\ref{eq:As1}) and (\ref{eq:At1}) should be evaluated at the horizon crossing, i.e., at $V=V_\star$.

The amplitude of the density perturbations fixes the brane tension. Taking the high-energy limit of Eq.~(\ref{eq:As1}) one can write
\begin{align}
\lambda \simeq \left[\frac{2\, \alpha^6}{75 \pi^2} \left(1+N_\star\right)^4\right]^{-1} A_s^2 ~.\label{eq:tensionAs}
\end{align}

The spectral tilt for scalar perturbations can be written in terms of the
slow-roll parameters as
\begin{equation}
n_{s} - 1   \equiv {d\ln A_{s}^2 \over d\ln k} \simeq -6\,\epsilon_\star +
2\,\eta_\star~, \label{eq:ns}
\end{equation}
while the tensor power spectrum can be parameterized in terms of the tensor-to-scalar ratio as
\begin{equation}
r_s\equiv 16 \,\dfrac{A_t^2}{ A_s^2}~,\label{eq:rs}
\end{equation}
which in the high-energy limit leads to
\begin{equation}
r_s \simeq 24 \,\epsilon_\star~.\label{eq:rs2}
\end{equation}

From Eqs.~(\ref{eq:epsiloneta})-(\ref{eq:Vstar}), the spectral index of the density perturbations and the tensor to scalar are found to be
\begin{align}
n_s&=1-\frac{4}{N_\star+1}\,,\label{eq:nsNstar}\\
r_s&= \frac {24}{N_\star+1}\,.\label{eq:rsNstar}
\end{align}
These results are both independent of the potential parameter $\alpha$ and the brane tension $\lambda$.

Finally, the running of the scalar spectral index $\alpha_s$ can be written as
\begin{equation}
\alpha_s  \equiv \frac{d\,n_s}{d \ln k} \simeq 16\, \epsilon_\star \eta_\star - 18\, \epsilon_\star^2 - 2\, \xi_\star ~,\label{eq:alphas}
\end{equation}
in the high-energy limit, where
\begin{equation}
\xi \simeq \frac{4\, \lambda^2 \, V^{\prime} V^{\prime \prime \prime} }{V^4}
\end{equation}
is the ``jerk" parameter.

\subsection{Observational constraints}

\begin{figure}[t]
\includegraphics[width=8cm]{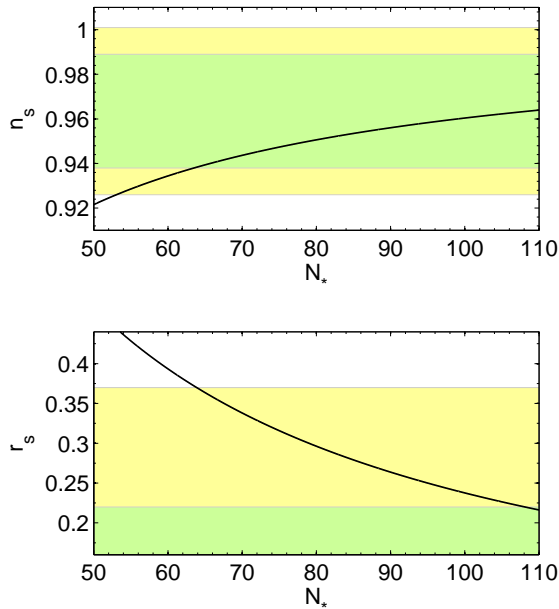}
\caption{The spectral index $n_s$ and tensor-to-scalar ratio $r_s$ as functions of the number of $e$-folds $N_\star$, Eqs. (\ref{eq:nsNstar}) and (\ref{eq:rsNstar}). The shaded areas are the bounds given in Eqs.~(\ref{eq:nsbound}) and (\ref{eq:rsbound}).} \label{fig:nsrs}
\end{figure}

The recent publication of the 5-year results of WMAP~\cite{Hinshaw:2008kr,Dunkley:2008ie,Komatsu:2008hk} puts very accurate constraints on the spectral index: $n_s = 0.963^{+0.014}_{-0.015}$ at $68\%$ confidence level (C.L.), for vanishing running and no tensor modes. This value is slightly above the 3-year result, $n_s = 0.958 \pm 0.016$~\cite{Spergel:2006hy}, and has a smaller uncertainty.

In what concerns the tensor modes, WMAP5~\cite{Komatsu:2008hk} alone gives $r_s < 0.43$ (with vanishing running) and $r_s < 0.58$ (with running), both at $95\%$ C.L. However, the strongest overall constraint on the tensor mode contribution comes from the combination of CMB, large-scale structure measurements and supernovae data. The combination of WMAP5, baryon acoustic oscillations (BAO) in the distribution of galaxies and supernovae~\cite{Komatsu:2008hk} yields $r_s < 0.20$ (without running) and $r_s < 0.54$ (with running), at $95\%$ C.L.

Since the running $\alpha_s$ is very small in the model under consideration, $|\alpha_s| \sim {\cal O} (10^{-3}- 10^{-4})$, we can make use of the observational bounds obtained for the case of vanishing running. However, the tensor modes cannot be neglected in this model. The inclusion of tensor modes has implications for $n_s$ as well: the constraint becomes $n_s = 0.968 \pm 0.015$ at $68\%$ C.L., for combined WMAP5, BAO and supernovae data.

In our analysis we shall consider the $99.9\%$ C.L. bounds obtained in  Ref.~\cite{Seljak:2006bg}, which take into account the 3-year results of WMAP together with other CMB experiments, galaxy surveys and supernovae data, as well as the Lyman-$\alpha$ forest power spectrum data from the Sloan Digital Sky Survey (SDSS). These bounds are~\cite{Seljak:2006bg}
\begin{align}
n_s & = 0.964 \,^{+0.025}_{-0.024} ~ \left(\,^{+0.037}_{-0.038}\,\right)~,\label{eq:nsbound}\\
r_s & < 0.22~ \left(\,<0.37\,\right)~.\label{eq:rsbound}
\end{align}
The error bars are at $2\sigma$ ($3\sigma$) and the upper bounds at $95\%$ ($99.9\%$) C.L.

In order to comply with the bounds of Eqs.~(\ref{eq:nsbound}) and (\ref{eq:rsbound}), one needs $N_\star \gtrsim 63 \,(53)$ and $N_\star \gtrsim 108 \,(64)$, respectively, [cf. Equations~(\ref{eq:nsNstar}) and (\ref{eq:rsNstar}); see also Fig.~\ref{fig:nsrs}]. Hereafter we use the $99.9\%$ C.L. lower bound coming from $r_s$,
\begin{align}
N_\star \gtrsim 64~.\label{eq:Nstarbound}
\end{align}

\begin{figure}[h]
\includegraphics[width=7cm]{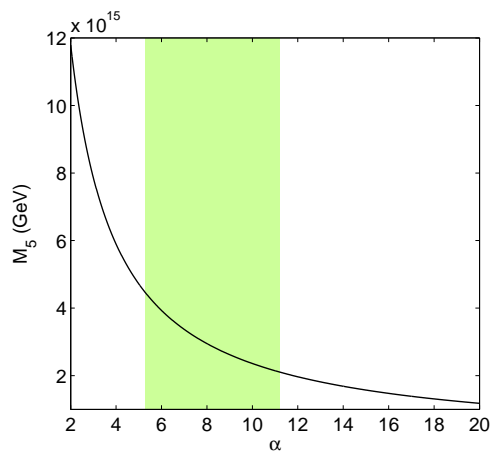}
\caption{The fundamental 5D Planck mass as a function of the parameter $\alpha$ [see Eq.~(\ref{eq:M5})] for $N_\star=64$. The shaded area corresponds to the values of $\alpha$ allowed by the observational constraints.} \label{fig:M5}
\end{figure}

The brane tension $\lambda$ can be determined from Eq.~(\ref{eq:tensionAs}) by imposing the correct amplitude for the density fluctuations, as measured by the WMAP team: $A_s^2 (k=0.002~ {\rm Mpc}^{-1}) \approx 4 \times 10^{-10}$~\cite{Spergel:2006hy,Komatsu:2008hk}. One finds
\begin{equation}
\lambda \simeq \frac {1.5 \times 10^{-7}}{\alpha^6} \left(\frac{M_4} {N_\star+1}\right)^4~,
\end{equation}
or, in terms of the 5D fundamental Planck mass,
\begin{equation} \label{eq:M5}
M_5 \simeq \frac{1.6\times 10^{-1}}{\alpha \,(N_\star+1)^{2/3}} \,M_4 \lesssim 1.9 \times 10^{-3} M_4~,
\end{equation}
where we have recovered the 4D Planck mass to help in noticing that this mass is of the order of the typical unification scale in grand unified theories. To obtain the upper bound we have used the lower bounds $\alpha \gtrsim 5.3$ (see Sec.~\ref{sec:quintessence}) and $N_\star \gtrsim 64$ (cf. Figure \ref{fig:M5}).

\section{From inflation to quintessence}

Braneworld inflation induced by the steep exponential potential ends when the inflaton field $\phi$ takes the value
\begin{equation}
\phi=\phi_{\rm end} \simeq \frac{1}{\alpha} \ln{\frac{V_0}{V_{\rm end}}}\,,
\label{phiend}
\end{equation}
where $V_0= A+\phi_0^2\,$, for the AS potential with $\phi_{\rm end} \ll \phi_0$.

The kinetic energy of the field at the end of inflation can also be easily estimated. Indeed, during the slow-roll period one has $3 H \dot{\phi} \simeq - V^\prime$, which in the high-energy limit of brane cosmology leads to $\dot{\phi} \simeq \alpha \sqrt{2\, \lambda/3}$. Therefore, taking into account Eq.~(\ref{eq:Vend}) we find
\begin{equation}
\dot{\phi}_{\rm end} = \sqrt{\frac{V_{\rm end}}{3}}~.\label{eq:phidotend}
\end{equation}

At the end of inflation, the Universe is in a cold and low-entropy state and it must be reheated to become a high-entropy and radiation-dominated Universe. Such a reheating process could occur, for instance, through the coherent oscillations of the inflaton field about the minimum of the potential until the age of the Universe equals the lifetime of the inflaton. The latter decays into ordinary particles, which then scatter and thermalize. However, our scenario of quintessential inflation belongs to the class of nonoscillatory models where the conventional reheating mechanism does not work: there is no minimum near $\phi_{\rm end}$ and the inflaton field cannot decay.

Therefore, reheating should be achieved by other means. One possibility is to assume that the Universe was reheated by the gravitational particle production at the end of the inflationary period~\cite{Ford:1986sy,Spokoiny:1993kt,Grishchuk:1990bj}. This is a democratic process which leads to the production of a variety of species quantum mechanically by the changing gravitational field at the end of inflation. Unlike the conventional reheating mechanism, this process does not require the introduction of extra fields. The radiation density created via this mechanism is given by
\begin{equation}
\rho_r \sim 0.01 \, g_p \, H_{\rm end}^4\,,
\label{eq:gravreh}
\end{equation}
where $g_p$ is the number of different particle species created from vacuum, likely to be ${\mathcal O}(10) \lesssim g_p \lesssim {\mathcal O}(100)$. Using Eqs.~(\ref{eq:Vend}), (\ref{eq:tensionAs}) and (\ref{eq:gravreh}), it can be easily shown that
\begin{equation}
\frac{\rho_\phi^{\rm end}}{\rho_r^{\rm end}} \sim 3.2 \times 10^{9}\left(N_\star + 1\right)^4 g_p^{-1}\,.
\label{eq:gravreh_ratio}
\end{equation}
For $N_\star \gtrsim 64$ one gets $\rho_r / \rho_\phi \lesssim 10^{-17} g_p$, which implies that the equality between the scalar field and radiation energy densities is reached very late. This leads to a prolonged kinetic regime during which $\rho_\phi \gg \rho_r$ and $p_\phi \simeq \rho_\phi$. It can be shown~\cite{Sahni:2001qp} that such a prolonged regime with a ``stiff" equation of state will generate an excessive gravity wave background which violates the BBN bound.

Alternatively, the Universe could have been instantaneously reheated via the so-called instant preheating mechanism~\cite{Felder:1998vq,Kofman:1994rk,Kofman:1997yn,Felder:1999pv}. Since this method turns out to be the most efficient in the context of quintessence models, it is assumed here as the reheating mechanism operative at the end of inflation.

\subsection{Braneworld inflation followed by instant preheating}

A successful reheating after inflation can be easily achieved in a field-theoretical framework where the inflaton $\phi$ interacts with another scalar field $\chi$ which, in turn, couples minimally to a fermionic field $\psi$ through a Yukawa coupling $h_f$~\cite{Felder:1998vq,Felder:1999pv}. The simplest interaction Lagrangian is
\begin{equation}
L_{\rm int}=-{1\over 2}g^2 \phi^2 \chi^2-h_f \bar{\psi}\psi \chi~.
\label{lagrangian}
\end{equation}
The process of $\chi$-particle production takes place as soon as $m_{\chi} = g\vert\phi\vert$ begins changing nonadiabatically~\cite{Felder:1998vq,Felder:1999pv}
\begin{equation}
\vert \dot{m}_\chi \vert \gtrsim m_\chi^2~.
\end{equation}
This condition is satisfied immediately after inflation has ended when
\begin{equation}
\vert \phi \vert \lesssim \vert \phi_{\rm prod} \vert = \sqrt{\frac{\vert \dot{\phi}_{\rm end}\vert}{g}}~,
\end{equation}
provided that $g\gg10^{-9}$~\cite{Sami:2004ic}.
The energy density of the created $\chi$-field particles is then given by
\begin{equation}
\rho_{\chi} =  m_\chi n_{\chi} \left({a_{\rm end} \over a}\right)^3~,
\end{equation}
where  $n_{\chi}= g^{3/2} (V_{\rm end}/2\pi)^3$ and the $({a_{\rm end} / a})^3$ term accounts for the cosmological dilution of the energy density with time.

If the quanta of the $\chi$-field are converted (thermalized)
into radiation instantaneously, the radiation energy density becomes~\cite{Felder:1998vq}
\begin{equation}
\rho_r \simeq \rho_\chi \sim  \left(\frac{g^{1/2}
V_{\rm end}}{2\pi}\right)^{3} g \, \phi_{\rm prod} \sim 10^{-2} g^2 V_{\rm end}~.
\label{irad}
\end{equation}
Thus, at the time inflation ends,
\begin{equation}
{\rho_r^{\rm end}  \over \rho_\phi^{\rm end}}  \sim 10^{-2} g^2~.
\label{irad1}
\end{equation}
Clearly, the energy density created by instant preheating can be much larger than the energy density produced by quantum particle production, for which $\rho_r/\rho_\phi \simeq 10^{-17} g_p$. In Ref.~\cite{Sami:2004xk} it was shown that in order to evade the BBN constraint on the energy density of relic gravity waves $\rho_{g}$ at the start of the radiative era, $\rho_g^{\rm rad}/\rho_r^{\rm rad}\lesssim 0.2$, one must have $\rho_r^{\rm end}/\rho_\phi^{\rm end}\gtrsim 10^{-7}$. Therefore, from the relation (\ref{irad1}) we should have $g \gtrsim 10^{-3}$.

The reheating process occurs through the decays of $\chi$ particles into fermions, as a consequence of the interaction term in the Lagrangian (\ref{lagrangian}). One can show that there is a wide region in the parameter space $(g, h_f)$ for which reheating is rapid and the relic gravity background in nonoscillatory braneworld models of quintessential inflation is consistent with the nucleosynthesis constraints. In fact, it is possible to derive a lower bound on $h_f$ so that the decay of the $\chi$-particles is sufficiently rapid and the reheating can be considered instantaneous: one should have $h_f \gtrsim 10^{-4} g^{-1/2}$~\cite{Sami:2004xk}.

\subsection{Initial conditions for quintessence}

In order to integrate the equations of motion (\ref{eq:Friedmann}) and (\ref{eq:kg}), it is convenient to rewrite them in the form~\cite{Copeland:1997et,Huey:2001ae,Huey:2001ah}
\begin{eqnarray}
{d x \over d \cal{N}} & = &-3 x + \sigma \sqrt{3\over 2}\,y^2+{3\over 2}\, x \,[2 x^2+\gamma(1-x^2-y^2)]~,\nonumber\\
{d y \over d \cal{N}}  & = &-\sigma\sqrt{3\over 2}\, x\, y + {3\over 2}\, y\, [2 x^2+\gamma(1-x^2-y^2)],
\label{eq:system}
\end{eqnarray}
where $\gamma =w_B+1$, ${\cal N} \equiv \ln a$, and
\begin{eqnarray}
x &\equiv &{\dot{\phi} \over \sqrt{2\,\rho}}~,\quad
y \equiv   {\sqrt{V} \over \sqrt{\rho}}~,\\
\sigma &\equiv & -  {V^\prime\over V} \left(1+{\rho \over 2 \lambda}\right)^{-1/2} ~.
\label{eq:defs}
\end{eqnarray}
We have to fix the initial conditions for the different energy components: scalar field, radiation, and matter densities. For the scalar field we assume that the initial value is given by the field value at the end of inflation, i.e.,
\begin{align}
\phi_i=\phi_{\rm end}~,\quad \dot{\phi}_i=\dot{\phi}_{\rm end}~.
\end{align}

To set the initial value for the radiation component we use Eq.~(\ref{irad1}), obtained from the instant preheating mechanism discussed above. We have then
\begin{equation}
\rho_r^i = 10^{-2} g^2 \rho_\phi^{\rm end}~,\label{eq:rhoradi}
\end{equation}
where we leave $g$ as a free parameter to be constrained by observations.

The beginning of the integration is fixed by imposing the correct amount of radiation $\rho_r^0$ at present,
\begin{equation}
1+z_i= \left(\frac{\rho_r^i}{\rho_r^0}\right)^{1/4}~.
\end{equation}
For the fraction of radiation at present we use the central value $\Omega_r^0 \,h^2 = 4.3 \times 10^{-5}$, which assumes the addition of three neutrino species. The matter content at $z_i$ is such that the correct fraction is reproduced at present, $\Omega_m^0 \,h^2 = 0.1369\pm 0.0037\,$~\cite{Komatsu:2008hk}. In our computations we will allow this quantity to vary in this interval.

\section{Quintessence}
\label{sec:quintessence}

\begin{figure}[t]
\includegraphics[width=8cm]{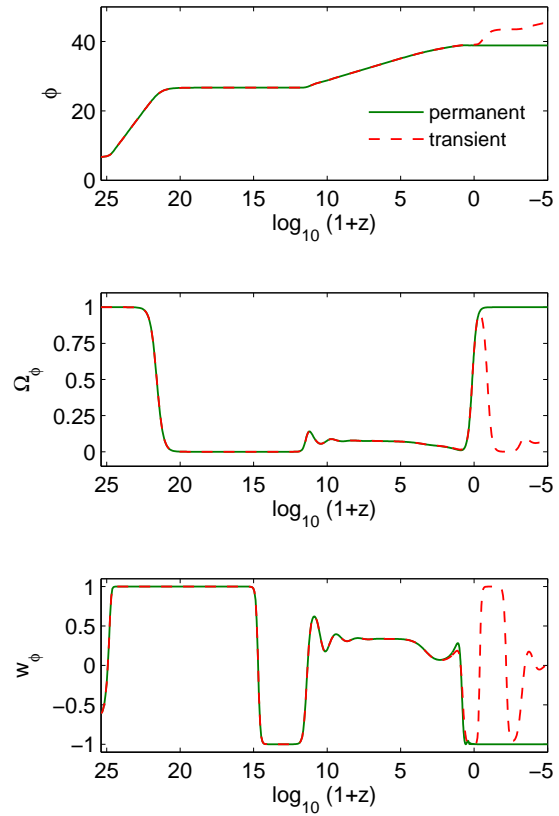}
\caption{Evolution of the scalar field, fractional energy density and equation of state of dark energy (from top to bottom) for $g= 0.025$, $\Omega_m^0\,h^2=0.137$, and two sets of the potential parameters $(\alpha,\,A,\,\phi_0)= (7,\,0.01,\,38.85)$ and $(7,\,0.02,\,38.925)$, leading to permanent (solid lines) and transient (dashed lines) acceleration, respectively.} \label{fig:evolution}
\end{figure}

As discussed above, the scalar field with exponential potential  leads to a viable evolution at early times. We should, however, ensure that the scalar field becomes quintessence at late times. In fact, any scalar field potential which interpolates between an exponential at early epochs and a power-law type potential at late times could lead to a viable cosmological evolution. The AS potential in Eq.~(\ref{eq:albrecht}) provides such an example. For illustration, in Fig.~\ref{fig:evolution}, we show the evolution of the scalar field, the fractional energy density, and equation of state of dark energy for two sets of parameters, one leading to permanent acceleration and the other to transient acceleration.

\subsection{Post-inflationary evolution}

\begin{figure*}[t]
\includegraphics[width=16cm]{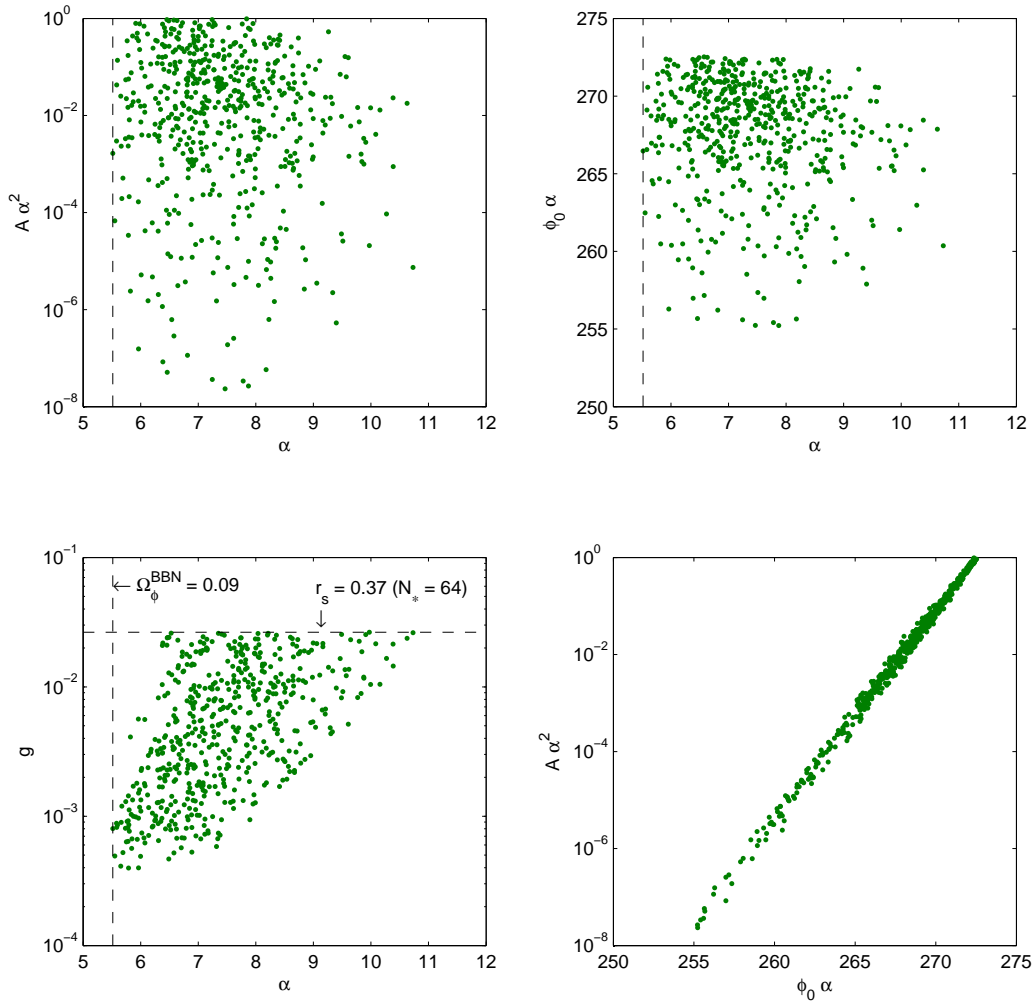}
\caption{Parameter space consistent with all the observational constraints considered, for the permanent acceleration case.} \label{fig:permanent}
\end{figure*}

\begin{figure*}[t]
\includegraphics[width=16cm]{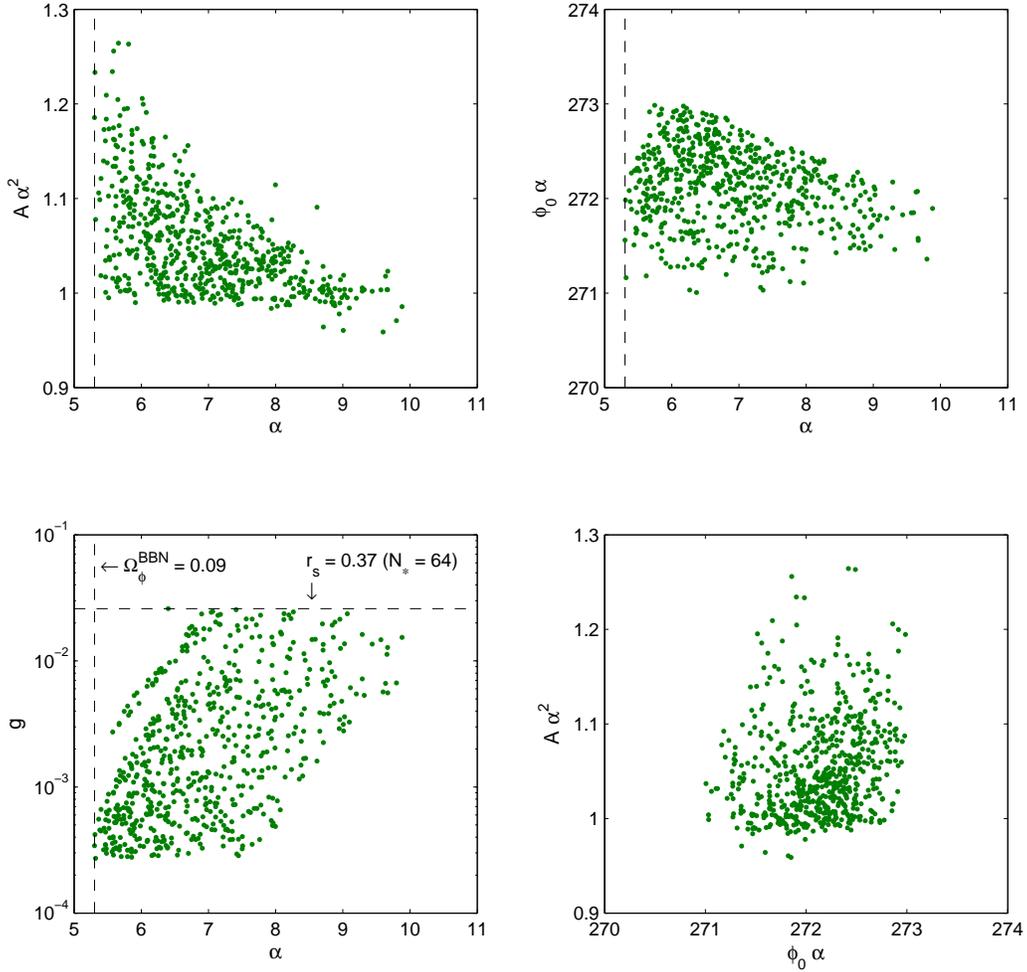}
\caption{Same as in Fig. \ref{fig:permanent}, but for the transient acceleration case.} \label{fig:transient}
\end{figure*}

After inflation ends, it takes a little while for the brane corrections to disappear and for the kinetic regime to commence. When $\rho \simeq \rho_\phi \lesssim 2 \lambda$, the scalar field rolling down a steep potential is now subject to minimum damping and soon goes into a free fall mode during which $\dot{\phi}^2 \gg V(\phi)$ and $\rho_\phi \propto a^{-6}$. The Universe undergoes a transition from an era dominated by the scalar field potential energy to a kination era. Since during the kination epoch $\rho_r/\rho_\phi \sim a^{2}$, the Universe eventually makes a transition to the standard radiation era (cf. Fig.~\ref{fig:evolution}).

We should mention that the number of $e$-folds from horizon crossing to the end of inflation, $N_\star$, can now be determined from the following considerations. A length scale $k_\star$ which crosses the Hubble radius during the inflationary epoch ($a_\star$) and reenters it today ($a_0$) will satisfy $k_\star= a_\star H_\star = a_0 H_0$, or equivalently,
\begin{equation}
\frac{k_\star}{a_0 H_0}=\frac{a_\star H_\star}{a_0 H_0}= \frac{a_\star}{a_{\rm end}} \frac{a_{\rm end}}{a_{\rm eq}}  \frac{H_\star}{H_{\rm eq}} \frac{a_{\rm eq} H_{\rm eq}}{a_0 H_0}\,,
\end{equation}
where $H_{\rm eq} = 4.4 \times 10^{-54}\, (\Omega_m^{0} \, h^2)^2 $ and $a_{\rm eq}$ are the values of the Hubble radius and the scale factor at the matter-radiation equality epoch, respectively; $H_\star$ is the Hubble radius at the horizon crossing scale $k_\star$,
\begin{equation}
H_\star^2 \simeq \frac{1}{3} \frac{V_\star^2}{2\lambda}\,.
\end{equation}
Taking into account that $a_\star/a_{\rm end}= e^{-N_\star}$ and
\begin{equation}
\frac{a_{\rm eq} H_{\rm eq}}{a_0 H_0} = 217.7\, \Omega_m^0 \, h\,,
\end{equation}
one can write
\begin{align}
N_\star =& \ln \frac{k_\star^{-1}}{3000\, h^{-1} \mbox{Mpc}} +
\ln{\frac{a_{\rm end}}{a_{\rm eq}}}\nonumber\\
&+\ln  \frac{V_\star}{\sqrt{6\lambda}\,H_{\rm eq}}
+ \ln 217.7\, \Omega_m^0\, h\,.
\label{eq:Nstar_computation}
\end{align}
To estimate $a_{\rm end}/a_{\rm eq}\,$, one can track the radiation evolution, $\rho_r \propto a^{-4}$. Using relation (\ref{eq:rhoradi}) we find
\begin{equation}
\frac{a_{\rm end}}{a_{\rm eq}}= \left( \frac{g}{10}\right)^{-1/2} \left( \frac{\rho_{\rm eq}}{\rho_\phi^{\rm end}}\right)^{1/4}~,
\end{equation}
where $\rho_{\rm eq} = 3 H_{\rm eq}^2$ denotes the energy density at matter-radiation equality epoch. Finally we obtain
\begin{align}
N_\star =& \ln \frac{k_\star^{-1}}{3000\, h^{-1} \mbox{Mpc}} +
\frac{1}{4}\ln{\frac{\rho_{\rm eq}}{\rho_\phi^{\rm end}}} - \frac{1}{2} \ln{\frac{g}{10}}\nonumber\\
&+\ln \frac{V_\star}{ \sqrt{6\lambda}\,H_{\rm eq}} + \ln 217.7\, \Omega_m^0\, h\,.
\label{eq:Nstar_computation}
\end{align}

Taking into account the lower bound on $N_\star$ given by Eq.~(\ref{eq:Nstarbound}), and noticing that $N_\star$ is almost independent of $\Omega_m^0\, h^2$ and $\alpha$, it is possible to derive an upper bound on the coupling $g$. Numerically we find that $g \lesssim 2.6\times10^{-2}$. This in turn implies a lower bound on the Yukawa coupling $h_f$. For the reheating of the Universe to be instantaneous at the end of the inflationary period, one should have $h_f \gtrsim 10^{-3}$.

\subsection{Late Time Evolution and observational constraints}

In our study, we perform a random analysis on the potential parameters $\alpha$, $A$, and $\phi_0$, together with $N_\star$ [$g$ is determined by Eq.~(\ref{eq:Nstar_computation})] and $\Omega_m^0 \,h^2 = 0.1369\pm 0.0037\,$. We consider the two possible late time behaviors: permanent or transient acceleration.

In order to have a viable model, besides the inflationary constraints, one needs to verify the other observational bounds on the different measured quantities during the different stages of the evolution of the Universe. As mentioned before, a stringent bound comes from the amount of dark energy during nucleosynthesis $\Omega_\phi^{\rm BBN} (z \simeq 10^{10}) \lesssim 0.09$~\cite{Bean:2001wt}. The bound arising from the CMB data, $\Omega_{\phi}^{\rm CMB}(z \simeq 1100)  < 0.39$~\cite{Bean:2001wt} at last scattering, is less stringent than the BBN bound. At present, we consider the following conservative bounds:
\begin{align}
0.6 \leq  h \leq 0.8~, \quad
0.6 \leq  \Omega_\phi^0 \leq 0.8~,\nonumber\\
w_\phi^0 \leq -0.8~,\quad
 q_0 < 0~,
\end{align}
where $q \equiv - \ddot{a}/(a\,H^2)$ is the deceleration parameter.

The results of our analysis\footnote{Notice that instead of $A$ and $\phi_0$ we use the combination of parameters $A \alpha^2$ and $\phi_0 \alpha$ to present our results. As already explained, $A \alpha^2$ determines the presence or absence of the minimum in the AS potential and, hence, it is useful for distinguishing between the permanent and transient regimes. The combination  $\phi_0 \alpha$ determines the position of the minimum/maximum or inflection point of the potential ($\phi_\pm = (1+ \phi_0 \alpha \pm \sqrt{1-A \alpha^2})/\alpha$), which is related to the exit from the tracking regime and to the scalar field energy density domination at present.} are displayed in Figs.~\ref{fig:permanent} and \ref{fig:transient}. We can see that the evolution of the scalar field $\phi$ from inflation to the present epoch is consistent with the observational constraints in a wide region of the parameter space of the AS potential, making it possible to obtain solutions with either eternal or transient accelerations.

The already discussed upper bound on the coupling $g$, coming from the $r_s$ constraint on $N_\star$, as well as the lower bound on the potential parameter $\alpha$, resulting from the bound on the amount of dark energy during BBN, is also shown in the figures (horizontal and vertical dashed lines, respectively). Decreasing $g$ or increasing $\alpha$ prolongs the kinetic regime. If this regime is too long, the history of the Universe is spoiled.  This allows us to put a lower and an upper bound on $g$ and $\alpha$, respectively. From the complete numerical analysis we find that the model exhibits transient acceleration at late times for
\begin{align}\label{eq:boundtrans1}
5.3 &\lesssim \alpha \lesssim 9.9\,,\nonumber\\
2.7 \times 10^{-4} &\lesssim g \lesssim 2.6\times10^{-2}\,,
\end{align}
\begin{align}\label{eq:boundtrans2}
0.96 \lesssim &A \alpha^2 \lesssim 1.26~,\nonumber\\
271 \lesssim\, &\phi_0 \alpha \lesssim 273~,
\end{align}
while permanent acceleration is obtained for
\begin{align}\label{eq:boundperm1}
5.5 &\lesssim \alpha \lesssim 10.8\,,\nonumber\\
4.0 \times 10^{-4} &\lesssim g \lesssim 2.6\times10^{-2}\,,
\end{align}
\begin{align}\label{eq:boundperm2}
2.3\times 10^{-8} \lesssim &A \alpha^2 \lesssim 0.98~,\nonumber\\
255 \lesssim\, &\phi_0 \alpha \lesssim 273~.
\end{align}
We notice that the bound on $\alpha$ is slightly lower than the one determined from Eq.~(\ref{track}), $\alpha \gtrsim 6.7$. We recall that Eq.~(\ref{track}) is only valid if the field is in the tracking regime, and, in the present model, it is possible to have initial conditions such that the scalar field has not yet entered the tracking regime during BBN.

The number of $e$-folds from horizon crossing till the end of inflation $N_\star$ and the value for the 5D Planck mass are very constrained:
\begin{align}\label{eq:boundNstarf}
64 \lesssim N_\star \lesssim 66~\,,
\end{align}
and (cf. Figure~\ref{fig:M5})
\begin{align}\label{eq:boundM5f}
 9.0 \times 10^{-4} \lesssim \dfrac{M_5}{M_4} \lesssim 1.9 \times 10^{-3} ~,
\end{align}
which imposes strong constraints on the inflationary observables $n_s$ and $r_s\,$.

\section{Conclusions}

We have analyzed a simple model of quintessential inflation in the RSII braneworld context with a  modified exponential potential. One of the attractive features of the model is that it can lead to transient acceleration at late times. This is particularly welcome in string theoretical formulations in order to avoid the difficulties arising in the $S$-matrix construction at the asymptotic future in a de-Sitter space~\cite{Hellerman:2001yi,Fischler:2001yj,Witten:2001kn}. Assuming that the Universe was reheated via the instant preheating mechanism, we have shown that the evolution of the scalar field from inflation till the present epoch is consistent with the observations in a wide region of the parameter space. Requiring that the model meets various cosmological constraints at the different stages of the evolution, we were able to constrain tightly its parameters, as summarized in Eqs.~(\ref{eq:boundtrans1})-(\ref{eq:boundM5f}).

In view of the very constrained bounds we obtained from the inflationary period, it is useful to consider how we could circumvent them in a simple and natural way. For instance, theoretical predictions for the inflationary observables may be modified by the presence of fields that are heavier than the Hubble rate during inflation. In this case, the coupling of the inflaton field to such heavy fields introduce corrections~\cite{Bartolo:2007hx} which can be larger than the second-order contributions in the slow-roll parameters. Another way to change predictions for these observables is to consider the more general framework of Gauss-Bonnet gravity. In the presence of the Gauss-Bonnet term, the value of the spectral index is determined by the Gauss-Bonnet coupling parameter and the tension of the brane and is independent of the slope of the potential, thereby bringing the scenario in closer agreement with the most recent observations~\cite{Lidsey:2003sj,Tsujikawa:2004dm}.


\begin{acknowledgments}
N.M.C.S. acknowledges the support of the Funda\c{c}\~{a}o para a Ci\^{e}ncia e a Tecnologia (FCT, Portugal) under the Grant No. SFRH/BPD/36303/2007. This work was also partially supported by FCT through the Project No. POCTI/FIS/56093/2004.
\end{acknowledgments}


\end{document}